\documentclass{ptapap}
\usepackage{hyperref}

\author{Daniel R. Reese}[LESIA]
\author{Marc-Antoine Dupret}[ULG]
\author{Michel Rieutord}[IRAP1, IRAP2]
\affil[LESIA]{LESIA, Observatoire de Paris, PSL Research University, CNRS,
           Sorbonne Universit\'es, UPMC Univ. Paris 06, Univ. Paris Diderot,
           Sorbonne Paris Cit\'e, 5 place Jules Janssen, 92195 Meudon, France}
\affil[ULG]{Institut d'Astrophysique et G\'eophysique de l'Universit\'e de Li\`ege,
            All\'ee du 6 Ao{\^u}t 17, 4000 Li\`ege, Belgium}
\affil[IRAP1]{Universit{\'e} de Toulouse, UPS-OMP, IRAP, Toulouse, France}
\affil[IRAP2]{CNRS, IRAP, 14 avenue Edouard Belin, 31400 Toulouse, France}

\title{Mode identification in rapidly rotating stars from BRITE data}

\newcommand{\eg}{\textit{e.g.}}
\newcommand{\ie}{\textit{i.e.}}
\newcommand{\Teff}{T_{\mathrm{eff}}}
\newcommand{\dTeff}{\delta T_{\mathrm{eff}}}
\newcommand{\Fbol}{\mathcal{F}_{\mathrm{bol}}}
\newcommand{\dFbol}{\delta \mathcal{F}_{\mathrm{bol}}}
\renewcommand{\l}{\ell}
\newcommand{\OmegaK}{\Omega_{\mathrm{K}}}

\newcommand{\Req}{R_{\mathrm{eq}}}

\begin{document}

\maketitle

\begin{abstract}

Apart from recent progress in Gamma~Dor stars, identifying modes in rapidly
rotating stars is a formidable challenge due to the lack of simple, easily
identifiable frequency patterns. As a result, it is necessary to look to
observational methods for identifying modes. Two popular techniques are
spectroscopic mode identification based on line profile variations (LPVs) and
photometric mode identification based on amplitude ratios and phase differences
between multiple photometric bands. In this respect, the BRITE constellation is
particularly interesting as it provides space-based multi-colour photometry. 
The present contribution describes the latest developments in obtaining
theoretical predictions for amplitude ratios and phase differences for pulsation
modes in rapidly rotating stars. These developments are based on full 2D
non-adiabatic pulsation calculations, using models from the ESTER code, the only
code to treat in a self-consistent way the thermal equilibrium of rapidly
rotating stars. These predictions are then specifically applied to the BRITE
photometric bands to explore the prospects of identifying modes based on BRITE
observations.

\end{abstract}

\section{Introduction}

Rapidly rotating stars intervene in many areas of astrophysics. For instance,
the majority of massive and intermediate mass stars are rapid rotators
\citep[\eg][]{Zinnecker2007, Royer2009}. Primordial stars are also expected to
be rapid rotators due to their low metallicity and hence, opacity
\citep[\eg][]{Ekstrom2008}. Rapid rotation is also thought to play a key role in
the precursors to gamma-ray bursts \citep[\eg][]{Woosley2006}. Understanding
these stars would yield valuable information to these areas of astrophysics.

Rapid rotation introduces many new phenomena in stars. These include
centrifugal deformation, gravity darkening, baroclinic flows, and various forms
of turbulence and transport phenomena \citep[\eg][]{Maeder2009, Rieutord2016}. 
As a result, there are many uncertainties in the models and a need for further
observational constraints. Currently, one of the best ways of constraining
stellar structure is through asteroseismology. However, identifying pulsation
modes in rapidly rotating stars, \ie\ finding the correspondence between
observed and theoretical pulsations, is a formidable challenge
\citep[\eg][]{Goupil2005}. Indeed, the pulsation spectra lack simple frequency
patterns, and reliable predictions for mode amplitudes are not available given
that modes in these stars tend to be excited by the $\kappa$ mechanism. As a
result, it is necessary to apply mode identification techniques.

There are two main types of mode identification techniques. The first is based
on multi-colour photometry and involves looking at ratios between pulsation
amplitudes, and phase differences, in different photometric bands. These
signatures depend on the geometry of the pulsation mode but are independent of
the intrinsic mode amplitudes. With its two colours, the BRITE mission is an
ideal source of space-based multi-colour photometric observations of pulsating
stars. The second approach is based on spectroscopy and consists in looking at
how the profile of a given absorption line changes with time. These variations
are known as line profile variations (LPVs) and provide a very rich information
which can be complementary to that provided by the photometric approach. Both
of these techniques need to be adapted to rapid rotation.

In the following, we will focus on photometric mode identification techniques.
We will specifically look at results in the BRITE photometric bands and see up
to what extent mode identification may be constrained. The next section
describes the prerequisites for coming up with reliable predictions as well as
the calculations carried out. This is followed by various results which focus
on amplitude ratios, phase differences, and complex asteroseismology. A
discussion concludes these proceedings.

\section{Calculating mode visibilities}

In order to calculate reliable mode visibilities, it is necessary to carry out
2D pulsation calculations which fully take into account the effects of rapid
rotation, in order to correctly calculate the geometry of the modes. For
instance, low-degree acoustic modes become island modes at rapid rotation rates
\citep{Lignieres2008, Lignieres2009}.  These modes take on an elongated
structure which circumvents the equator and is characterised by a new set of
quantum numbers $(\tilde{n},\,\tilde{\l},\,m)$ \citep[\eg][]{Reese2008}. 
Calculating such modes as well as other modes present in rapidly rotating stars
requires the use of stellar models which fully take into
account stellar deformation. Furthermore, non-adiabatic pulsation calculations
are required in order to correctly calculate $\dTeff/\Teff$, the variations in
effective temperature, as these intervene in the intensity variations used to
calculate mode visibilities. In order to carry out such calculations, it is
necessary for the stellar model to respect the energy conservation equation. 
This means that the stellar model will be baroclinic, \ie\ isobars, isotherms,
and isochores will not coincide, and the rotation profile will be
non-conservative, \ie\ it will depend on both $s$, the distance to the rotation
axis, and on $z$, the vertical coordinate.

In the work presented here, we will use ESTER\footnote{Evolution STEllaire en
Rotation.} models \citep{EspinosaLara2013, Rieutord2016} as these are currently
the only rapidly rotating models which satisfy the energy equation locally. 
Non-adiabatic calculations will be carried out using the
TOP\footnote{Two-dimensional Oscillation Program.} pulsation code
\citep[\eg][]{Reese2009a, Reese2017b}. For stars in the mass range of
$\delta$~Scuti stars, our implementation of non-adiabatic calculations are not fully reliable. 
Accordingly, we will also use models from the SCF\footnote{Self-Consistent
Field.} code \citep{Jackson2005, MacGregor2007} along with adiabatic pulsation
calculations. Non-adiabatic effects will be approximated in the same way as is
done in \citet{Reese2017}, \ie\ using the pseudo non-adiabatic (PNA) approach (see
Table~\ref{tab:success_rates}).

\section{Multi-colour photometric mode identification}

\subsection{Amplitude ratios}

As was shown in \citet{Daszynska_Daszkiewicz2002} and \citet{Townsend2003b}
amplitude ratios depend on the azimuthal order, $m$, and the inclination, $i$,
in rotating stars, thereby complicating the task of mode identification.
Nonetheless, \citet{Reese2013} found similar amplitude ratios for modes with the
same $(\l,m)$ values but different radial orders, $n$, as expected from ray
theory \citep{Pasek2012}. Accordingly, \citet{Reese2017} proposed an alternate
mode identification strategy. This strategy involves choosing a reference mode,
then choosing the $N$ (typically 9) other modes with the most similar amplitude
ratios. When the reference mode happens to be an island mode, the other modes
also tend to be island modes with similar quantum numbers. The corresponding
frequencies would then follow patterns as expected from the asymptotic frequency
formula \citep[\eg][]{Lignieres2009, Reese2009a}. By repeating this procedure,
one can hope to group similar modes together into families and identify
recurrent frequency spacings as expected from the asymptotic formula. An open
question is whether this strategy still continues to work when using the 2
photometric bands from BRITE rather than the 7 bands from the Geneva photometric
system.

In Table~\ref{tab:success_rates}, we give the average success rates at finding
other island modes using the Geneva and BRITE photometric systems for different
stellar masses. The third column gives the success rate at identifying other
island modes if the reference mode is an island mode, whereas the fourth and
fifth columns give the success rates at finding island modes with the same
$(\l,\,|m|)$ and $(\tilde{\l},\,|m|)$ values, respectively. We recall that two
modes with the same $(\tilde{\l},\,|m|)$ values will not necessarily have the
same $(\l,\,|m|)$ values as one could be symmetric with respect to the equator
and the other anti-symmetric. The last column gives the proportion of island
modes in the entire set of modes. As can be seen, the success rates for the
BRITE photometric system are much lower. Typically, one can expect to identify
1 or 2 other island modes out of a set of $n=9$ modes, which is insufficient for
the purposes of mode identification.

\begin{table}[tb]
\caption{Success rates for the mode identification strategy using the Geneva and
     BRITE photometric systems, for SCF models at $0.6\,\OmegaK$ (where
     $\OmegaK$ is the Keplerian break-up rotation rate).
     \label{tab:success_rates}}
\begin{center}
 \begin{tabular}{llcccc}
 \hline
 \hline
 & \textbf{Photo.} &
 \multicolumn{3}{c}{\dotfill \textbf{Success rates} \dotfill} &
 \textbf{Island} \\
 \textbf{Model} &
 \textbf{System} &
 \textbf{Island} &
 $(\l,\,|m|)$ &
 $(\tilde{\l},\,|m|)$ &
 \textbf{prop.} \\
 \hline
   Adia. (2\,M$_{\odot}$)   & Geneva &  0.564 &  0.359 &  0.416 &  0.0115 \\
   Adia. (2\,M$_{\odot}$)   & BRITE  &  0.145 &  0.058 &  0.071 &  0.0115 \\
   Adia. (1.8\,M$_{\odot}$) & Geneva &  0.554 &  0.401 &  0.452 &  0.0330 \\
   Adia. (1.8\,M$_{\odot}$) & BRITE  &  0.258 &  0.133 &  0.159 &  0.0330 \\
   PNA (1.8\,M$_{\odot}$)   & Geneva &  0.469 &  0.303 &  0.349 &  0.0330 \\
   PNA (1.8\,M$_{\odot}$)   & BRITE  &  0.201 &  0.079 &  0.102 &  0.0330 \\
  \hline
  \end{tabular} \\
  {\small PNA = pseudo non-adiabatic}
\end{center}
\end{table}

One may then wonder what happens if a supplementary photometric band is
included. For reasons of normalisation, we prefer not to mix visibilities from
the Geneva and BRITE systems. Hence, we use the B and G Geneva bands as
representative of the BRITE bands although we do note the latter are much
wider. Then we include either the U or V1 bands as these are centred around
the smallest and highest wavelengths, respectively, besides the B and G bands. 
Table~\ref{tab:success_rates_bis} gives the success rates for the BRITE
photometric system as well as reduced versions of the Geneva system. Columns 2
to 4 have the same meaning as columns 3 to 5 of Table~\ref{tab:success_rates}. 
As can be seen, adding one band, especially at small wavelengths, increases the
success rates appreciably.

\begin{table}[tb]
\caption{Success rates for the mode identification strategy for the BRITE and
     reduced versions of the Geneva photometric systems. These values are
     obtained for the $1.8$\,M$_{\odot}$ stellar model using pseudo
     non-adiabatic calculations.\label{tab:success_rates_bis}}
 \begin{center}
 \begin{tabular}{lccc}
 \hline
 \hline
 \textbf{Photo.} &
 \multicolumn{3}{c}{\dotfill \textbf{Success rates} \dotfill} \\
 \textbf{Bands} &
 \textbf{Island} &
 $(\l,\,|m|)$ &
 $(\tilde{\l},\,|m|)$ \\
 \hline
   BRITE    &  0.201 &  0.079 &  0.102 \\
   B, G     &  0.182 &  0.060 &  0.087 \\
   U, B, G  &  0.327 &  0.182 &  0.211 \\
   B, V1, G &  0.285 &  0.146 &  0.187 \\
  \hline
  \end{tabular}
  \end{center}
\end{table}

In summary, this approach is expected to start working for at least 3
photometric bands, and would also require a large number of acoustic modes,
preferably in the asymptotic regime. Hence, additional observations besides
those of BRITE are needed, and $\delta$~Scuti stars would be the most suitable
targets.

\subsection{Amplitude ratios and phase differences}

In some cases, nonetheless, observations may only be available in 2 rather than
3 bands. Furthermore, the number of observed modes may be too small for the
above strategy. This raises the question as to how much information can be
obtained from both amplitude ratios and phase differences. In what follows, we
use full non-adiabatic calculations as these are needed for obtaining reliable
phase differences. Accordingly, we will work with $9$\,M$_{\odot}$ ZAMS models
produced by the ESTER code,
with $X=0.700$, $Z=0.025$, and rotation rates ranging from $0.0$ to
$0.5\,\OmegaK$  (where $\OmegaK=\sqrt{GM/\Req^3}$ is the Keplerian break-up
rotation rate, $\Req$ being the equatorial radius).

\begin{figure}[tb]
\includegraphics[width=0.499\textwidth]{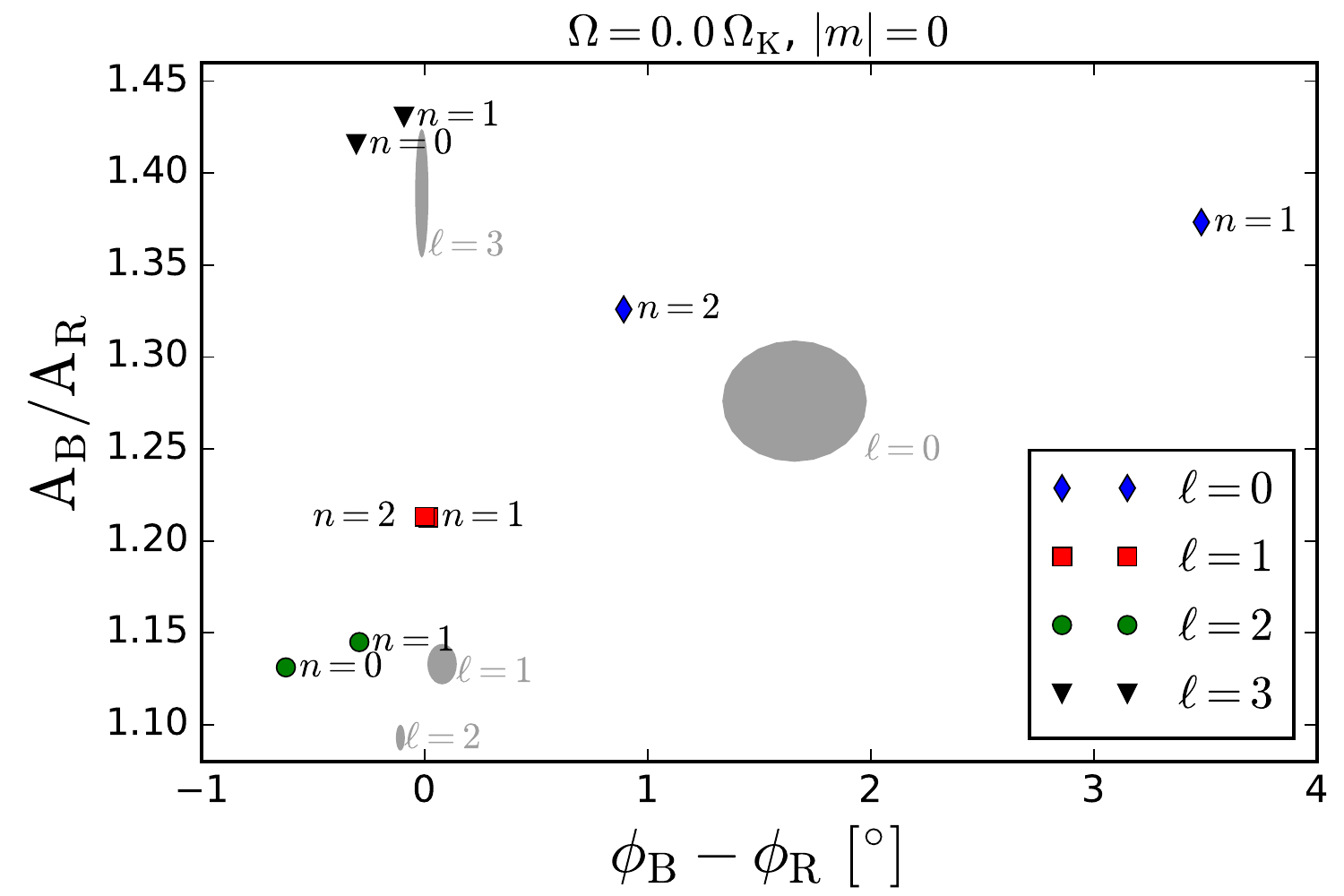} 
\includegraphics[width=0.499\textwidth]{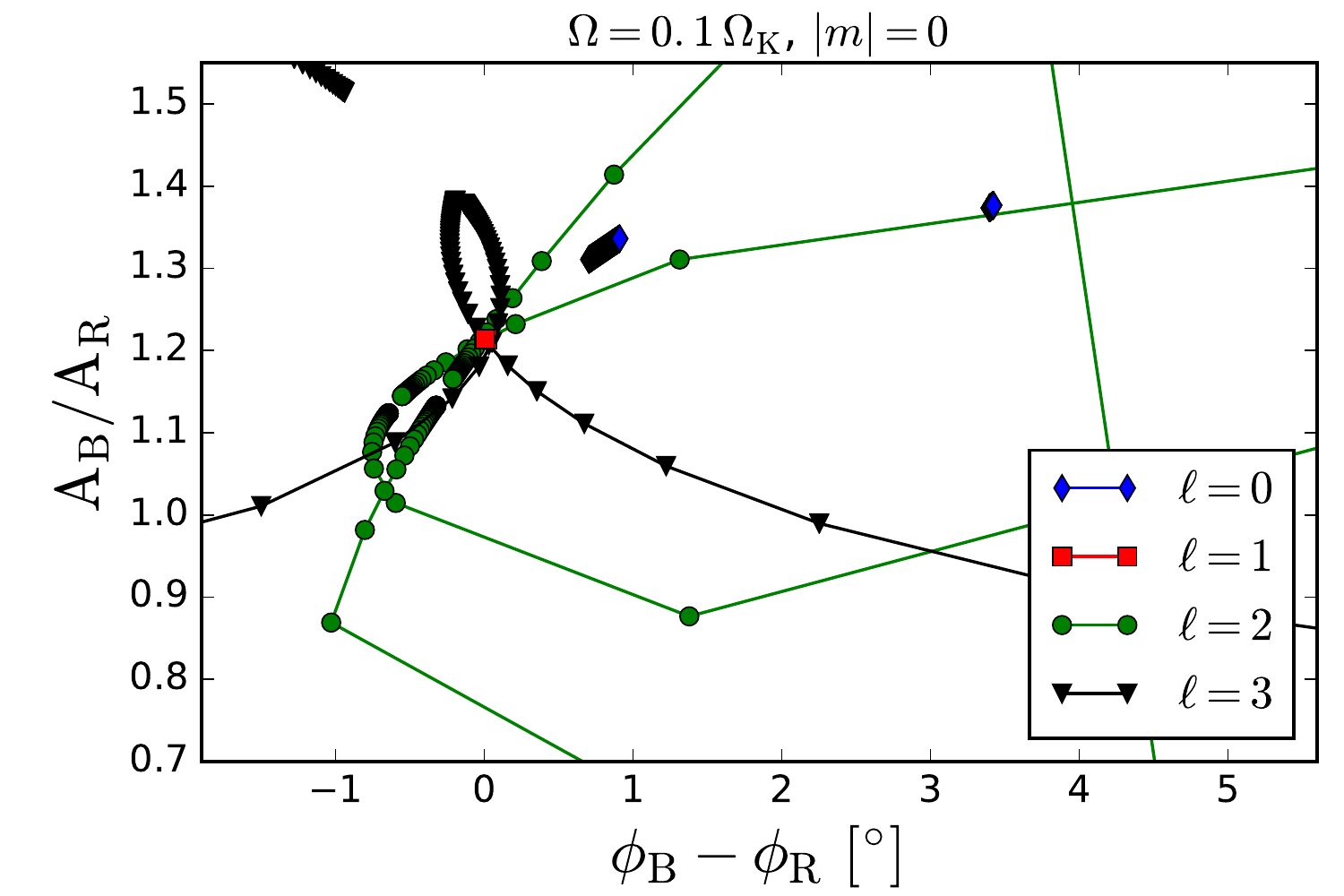} \\
\includegraphics[width=0.499\textwidth]{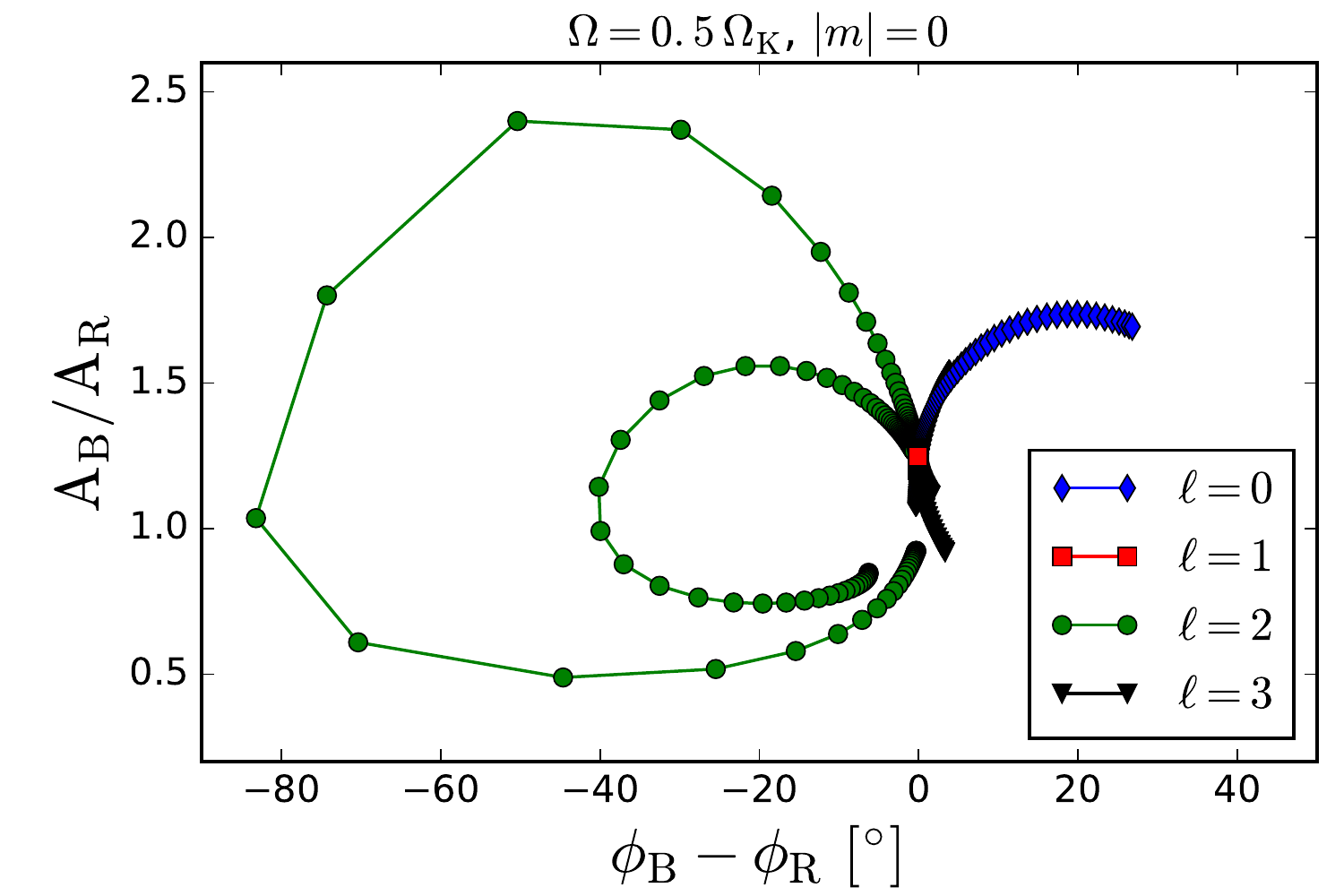} 
\includegraphics[width=0.499\textwidth]{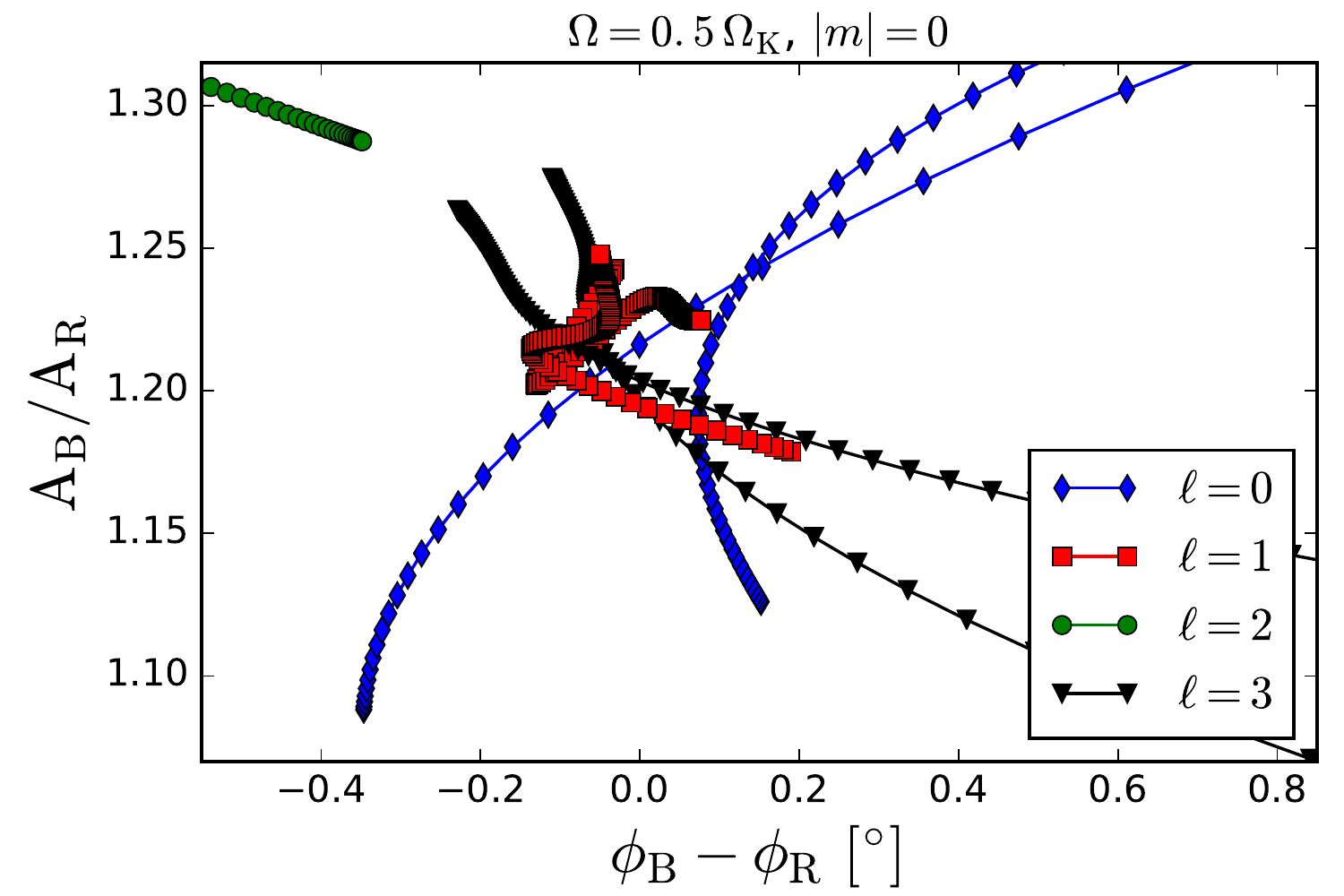} \\
\caption{Amplitude ratios vs. phase differences for axisymmetric modes in $9$\,M$_{\odot}$ ESTER models at different rotation rates using the BRITE photometric
bands. The lower right panel is a zoom in of the lower left panel. Lines
connect results for the same $(n,\,\ell)$ values but for stellar inclinations
ranging from $2^{\circ}$ to $89^{\circ}$ in increments of
$1^{\circ}$.\label{fig:ratios_phases}}
\end{figure}

As a first step, we compared our amplitude ratios vs.\ phase differences with
those from \citet[][hereafter H17]{Handler2017} for similar non-rotating models
in order to validate our calculations. The mass of our model is $9$\,M$_{\odot}$
whereas those of H17 range from $9.5$ to $10$\,M$_{\odot}$. The amplitude ratios
vs.\ phase differences are plotted in the upper left panel of
Fig.~\ref{fig:ratios_phases}, the grey regions corresponding to H17. As can be
seen, a qualitative agreement is obtained.

We then look at how rotation affects amplitude ratios and phases differences in
the remaining panels of Fig.~\ref{fig:ratios_phases}. As expected, these now
depend on the stellar inclination, in contrast to the non-rotating case. 
Furthermore, there are large excursions in these diagrams. This is typically
caused by mode amplitudes going to zero at slightly different stellar
inclinations in the different photometric bands. Accordingly, it seems unlikely
that such excursions will be seen in observed stars, since at least one of the
components will likely be below the detection threshold.

\begin{figure}[tb]
\includegraphics[width=0.499\textwidth]{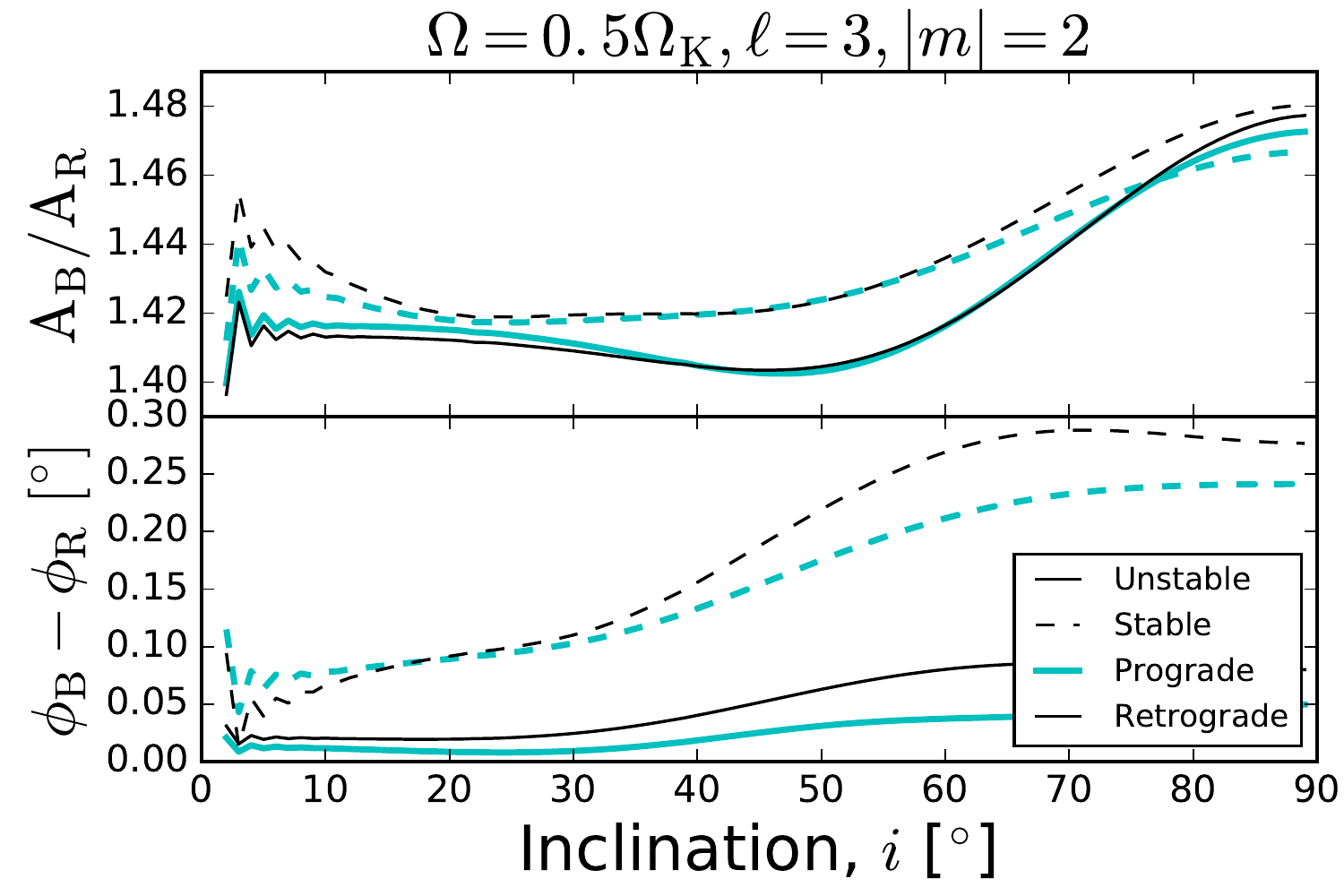}
\includegraphics[width=0.499\textwidth]{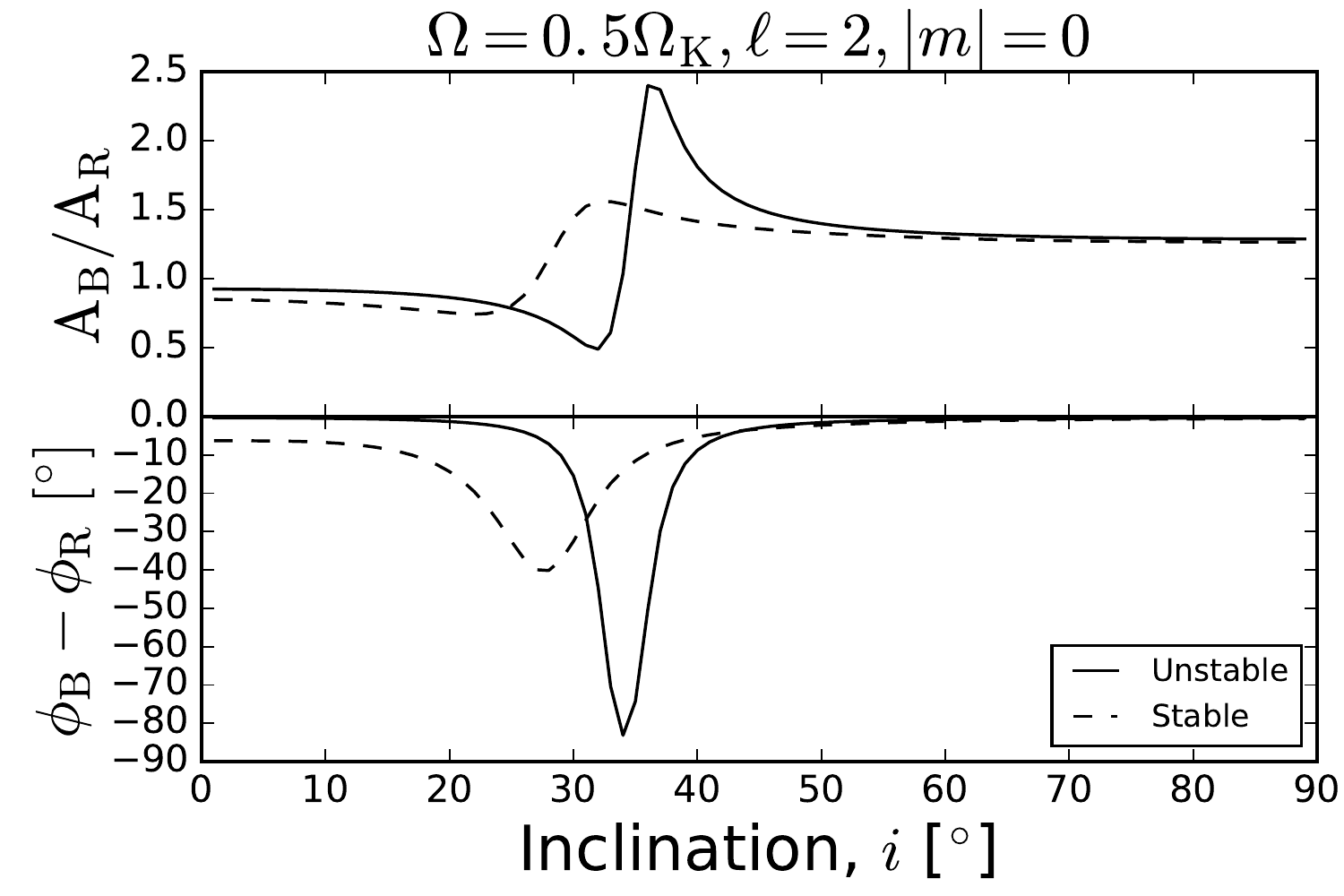} 
\caption{Amplitude ratios and phase differences as a function of inclination for
modes with the same $(\l,\,|m|)$ values. We note that the zigzags at low
inclinations in the left panel are numerical artifacts due to low mode
visibilities in both bands.\label{fig:ratios_phases_incl}}
\end{figure}

An important question is then whether amplitude ratios and phase differences
will be similar for modes with the same $(\l,\,|m|)$ values but different radial
orders, and whether this can help with mode identification.
Figure~\ref{fig:ratios_phases_incl} shows amplitude ratios and phase differences
for $(\l,\,|m|)=(3,2)$ in the left panel and $(2,0)$ in the right panel. As can
be seen, the left panel corresponds to a case where the amplitude ratios and
phases differences are similar, whereas the right panel shows a case with larger
differences, especially for $i\simeq 30^{\circ}$, due to large excursions at
different inclinations. Hence, the answer to the question depends both on the
choice of $(\l,\,|m|)$ and on the inclination. A more exhaustive search for a
large set of radial orders will be needed for $\delta$~Scuti type stars once
reliable non-adiabatic pulsation calculations are available for these.

Even with a limited number of modes, one can always apply a $\chi^2$
minimisation to find the modes which best reproduce observed amplitude ratios
and phases differences. In this regard, we recall the work done by
\citet{Daszynska_Daszkiewicz2015} in modelling the SPB star $\mu$~Eridani, where
such a minimisation was carried out in the framework of the traditional
approximation, only including excited modes, and taking into account the
observed $v\,\sin i$ value. This allowed them to constrain the mode
identification as well as the values of $v$ and $i$.

\subsection{Complex asteroseismology}

In \citeyear{Daszynska_Daszkiewicz2009}, \citeauthor{Daszynska_Daszkiewicz2009}
applied what they called ``complex asteroseismology'' to the $\beta$~Cephei star
$\theta$~Ophiuchi. Complex asteroseismology involves observationally determining
the $f$ parameter in addition to the mode identification, using both
multi-colour photometry and radial velocity measurements from spectroscopy,
where $f$ is the ratio of the bolometric flux perturbations to the radial
displacement:
\begin{equation}
\frac{\dFbol}{\Fbol} = 4 \frac{\dTeff}{\Teff} = f \frac{\xi_r}{R}.
\end{equation}
The $f$ parameter is complex due to non-adiabatic effects which introduce a
phase shift between the flux perturbations and radial displacements. This
parameter is independent of latitude in the non-rotating case because $\dFbol$
and $\xi_r$ are proportional to the same spherical harmonic. An open question
is what happens to $f$ when the star rotates rapidly.

\begin{figure}[tb]
\includegraphics[width=0.499\textwidth]{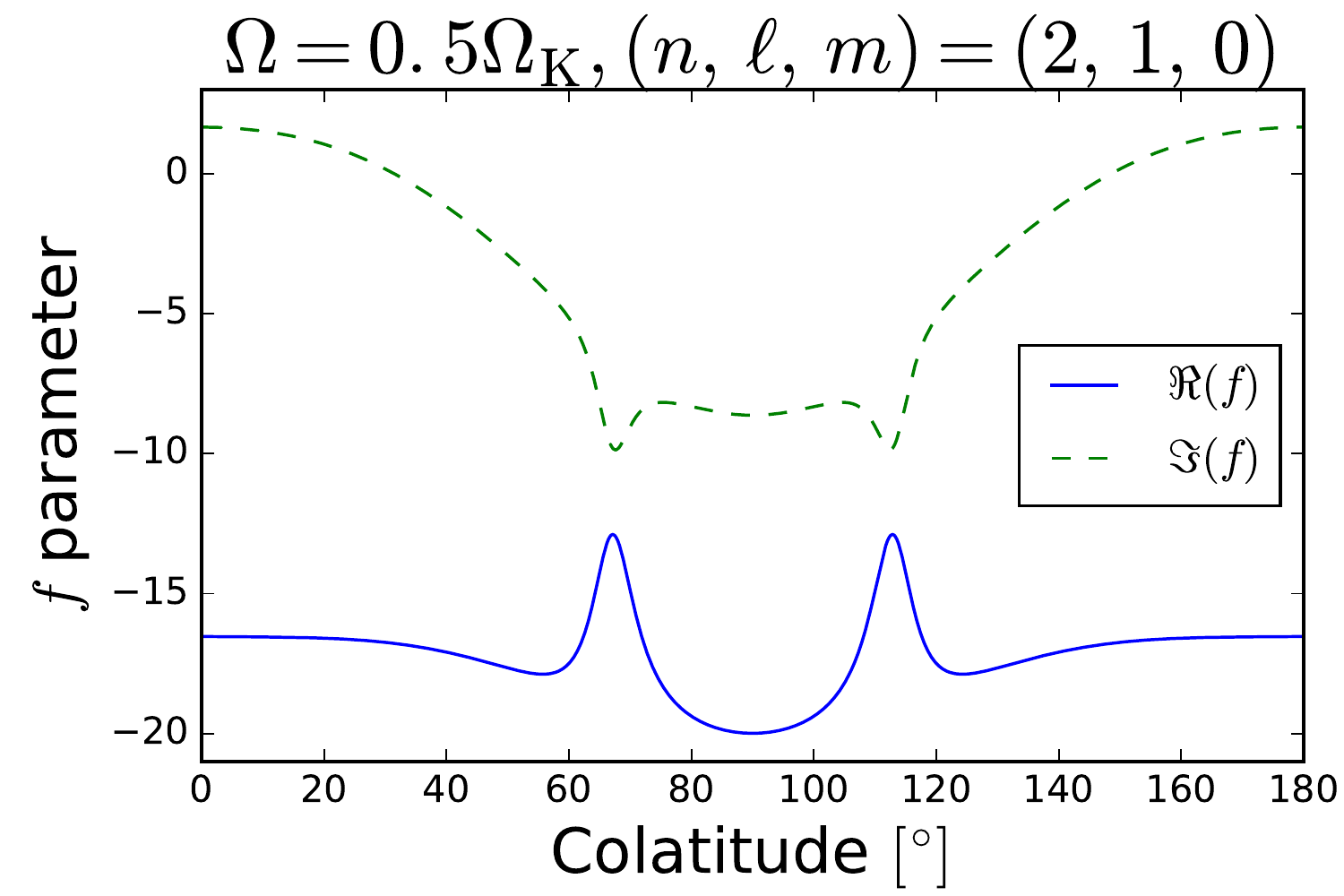}
\includegraphics[width=0.499\textwidth]{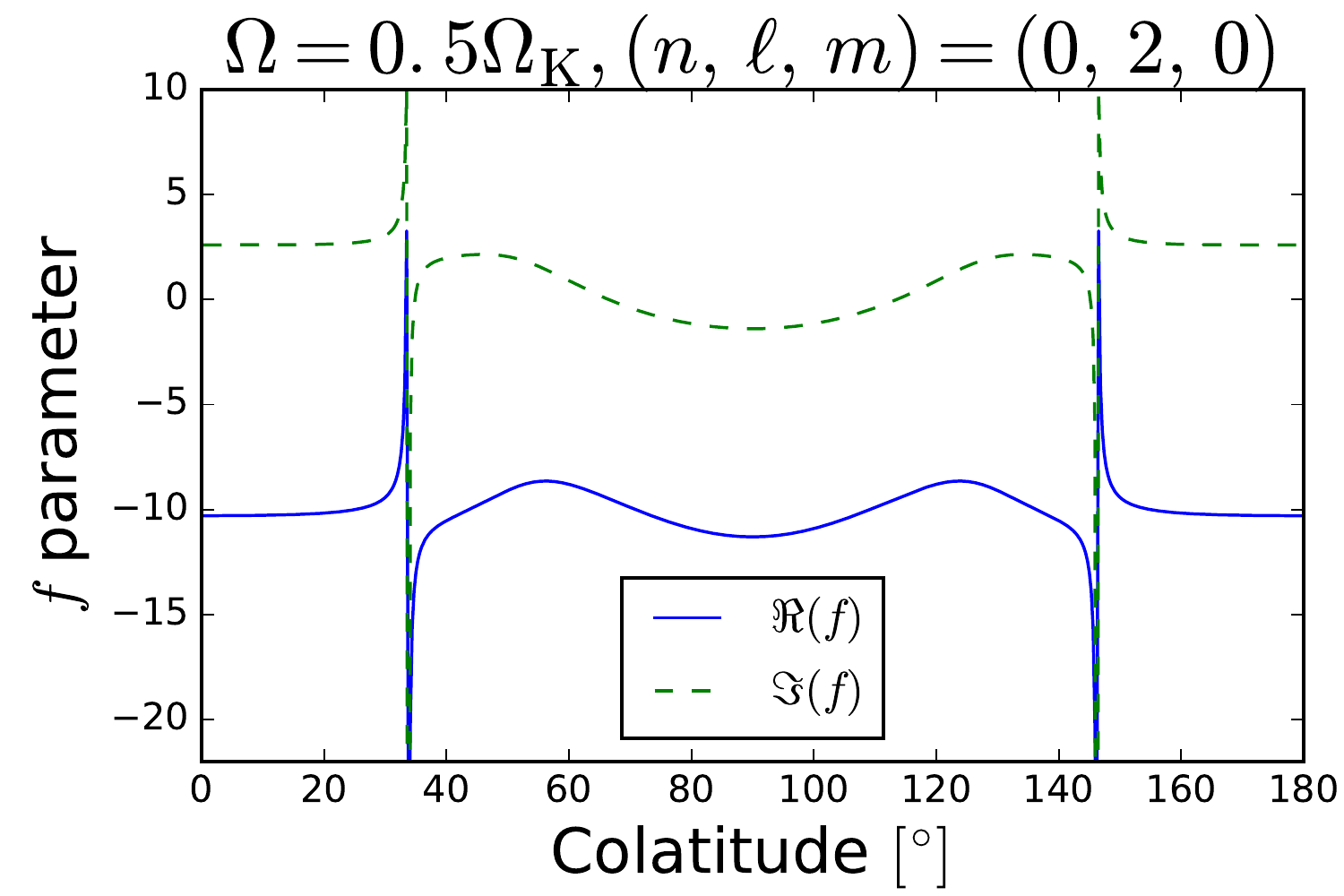} 
\caption{The $f$ parameter as a function of colatitude for two pulsation modes
in a rapidly rotating ESTER model. \label{fig:f_parameter}}
\end{figure}

In Fig.~\ref{fig:f_parameter}, we plot the real and imaginary parts of $f$ as a
function of colatitude for two different modes. In our definition of $f$, we
used the displacement perpendicular to the stellar surface, $\xi_{\mathrm{v}}$, 
normalised by the equatorial radius. As can be seen $f$ now depends on the
colatitude. Furthermore, sharp spikes occur in the right panel as a result of
$\dTeff$ and $\xi_{\mathrm{v}}$ going to zero at slightly different colatitudes.
Hence, the $f$ parameter should be described as an $f$ profile. This raises the
question as to whether it would be possible to define some sort of
disk-integrated, possibly inclination-dependent, effective $f$ parameter which
may still be used to constrain stellar structure, as is done in the non-rotating
case.

\section{Conclusion}

In summary, rotation leads to a much more complicated picture for amplitude
ratio and phase differences. It also leads to a latitude-dependant $f$
parameter, $f$ being the ratio between the bolometric flux perturbations and
radial displacement. Accordingly, this makes mode identification more difficult
and complicates complex asteroseismology. Conversely, this may provide tighter
constraints on the azimuthal orders and stellar inclination
\citep[\eg][]{Daszynska_Daszkiewicz2015}.

Different prospects include extending full non-adiabatic calculations to
$\delta$~Scuti stars, developing a database of mode visibilities and LPVs for
the purposes of identification, adapting and/or developing mode identification
tools, and fully exploiting pulsation data of rapidly rotating stars from the
BRITE mission as well as from PLATO 2.0 and the ground-based spectroscopic SONG
network.

\acknowledgements{
DRR and MR acknowledge the support of the French Agence Nationale de la
Recherche (ANR) to the ESRR project under grant ANR-16-CE31-0007. DRR also
acknowledges financial support from the ``Programme National de Physique
Stellaire'' (PNPS) of CNRS/INSU, France.
}

\bibliographystyle{ptapap}
\bibliography{biblio.bib}

\begin{thebibliography}{23}
\providecommand{\natexlab}[1]{#1}
\providecommand{\url}[1]{\texttt{#1}}
\providecommand{\urlprefix}{URL }
\providecommand{\eprint}[2][]{\url{#2}}

\bibitem[{{Daszy{\'n}ska-Daszkiewicz} et~al.(2015){Daszy{\'n}ska-Daszkiewicz},
  {Dziembowski}, {Jerzykiewicz}, \& {Handler}}]{Daszynska_Daszkiewicz2015}
{Daszy{\'n}ska-Daszkiewicz}, J., {Dziembowski}, W.~A., {Jerzykiewicz}, M.,
  {Handler}, G., \emph{{Oscillation modes in the rapidly rotating slowly
  pulsating B-type star {$\mu$} Eridani}}, \emph{MNRAS} \textbf{446}, 1438
  (2015)

\bibitem[{{Daszy{\'n}ska-Daszkiewicz} et~al.(2002){Daszy{\'n}ska-Daszkiewicz},
  {Dziembowski}, {Pamyatnykh}, \& {Goupil}}]{Daszynska_Daszkiewicz2002}
{Daszy{\'n}ska-Daszkiewicz}, J., {Dziembowski}, W.~A., {Pamyatnykh}, A.~A.,
  {Goupil}, M.-J., \emph{{Photometric amplitudes and phases of nonradial
  oscillation in rotating stars}}, \emph{A\&A} \textbf{392}, 151 (2002)

\bibitem[{{Daszy{\'n}ska-Daszkiewicz} \&
  {Walczak}(2009)}]{Daszynska_Daszkiewicz2009}
{Daszy{\'n}ska-Daszkiewicz}, J., {Walczak}, P., \emph{{Constraints on opacities
  from complex asteroseismology of B-type pulsators: the {$\beta$} Cephei star
  {$\theta$} Ophiuchi}}, \emph{MNRAS} \textbf{398}, 1961 (2009)

\bibitem[{{Ekstr{\"o}m} et~al.(2008)}]{Ekstrom2008}
{Ekstr{\"o}m}, S., et~al., \emph{{Effects of rotation on the evolution of
  primordial stars}}, \emph{A\&A} \textbf{489}, 685 (2008)

\bibitem[{{Espinosa Lara} \& {Rieutord}(2013)}]{EspinosaLara2013}
{Espinosa Lara}, F., {Rieutord}, M., \emph{{Self-consistent 2D models of
  fast-rotating early-type stars}}, \emph{A\&A} \textbf{552}, A35 (2013)

\bibitem[{{Goupil} et~al.(2005)}]{Goupil2005}
{Goupil}, M.-J., et~al., \emph{{Asteroseismology of {$\delta$} Scuti Stars:
  Problems and Prospects}}, \emph{JA\&A} \textbf{26}, 249 (2005)

\bibitem[{{Handler} et~al.(2017)}]{Handler2017}
{Handler}, G., et~al., \emph{{Combining BRITE and ground-based photometry for
  the {$\beta$} Cephei star {$\nu$} Eridani: impact on photometric pulsation
  mode identification and detection of several g modes}}, \emph{MNRAS}
  \textbf{464}, 2249 (2017)

\bibitem[{{Jackson} et~al.(2005){Jackson}, {MacGregor}, \&
  {Skumanich}}]{Jackson2005}
{Jackson}, S., {MacGregor}, K.~B., {Skumanich}, A., \emph{{On the Use of the
  Self-consistent-Field Method in the Construction of Models for Rapidly
  Rotating Main-Sequence Stars}}, \emph{ApJS} \textbf{156}, 245 (2005)

\bibitem[{{Ligni{\`e}res} \& {Georgeot}(2008)}]{Lignieres2008}
{Ligni{\`e}res}, F., {Georgeot}, B., \emph{{Wave chaos in rapidly rotating
  stars}}, \emph{Phys. Rev. E} \textbf{78}, 1, 016215 (2008)

\bibitem[{{Ligni{\`e}res} \& {Georgeot}(2009)}]{Lignieres2009}
{Ligni{\`e}res}, F., {Georgeot}, B., \emph{{Asymptotic analysis of
  high-frequency acoustic modes in rapidly rotating stars}}, \emph{A\&A}
  \textbf{500}, 1173 (2009)

\bibitem[{{MacGregor} et~al.(2007){MacGregor}, {Jackson}, {Skumanich}, \&
  {Metcalfe}}]{MacGregor2007}
{MacGregor}, K.~B., {Jackson}, S., {Skumanich}, A., {Metcalfe}, T.~S.,
  \emph{{On the Structure and Properties of Differentially Rotating,
  Main-Sequence Stars in the 1-2 M$_{\odot}$ Range}}, \emph{ApJ} \textbf{663},
  560 (2007)

\bibitem[{{Maeder}(2009)}]{Maeder2009}
{Maeder}, A., {Physics, Formation and Evolution of Rotating Stars}, {Astronomy
  and Astrophysics Library}, {Springer-Verlag} (2009)

\bibitem[{{Pasek} et~al.(2012){Pasek}, {Ligni{\`e}res}, {Georgeot}, \&
  {Reese}}]{Pasek2012}
{Pasek}, M., {Ligni{\`e}res}, F., {Georgeot}, B., {Reese}, D.~R.,
  \emph{{Regular oscillation sub-spectrum of rapidly rotating stars}},
  \emph{A\&A} \textbf{546}, A11 (2012)

\bibitem[{{Reese}(2008)}]{Reese2008}
{Reese}, D., \emph{{Modelling rapidly rotating stars}}, \emph{Journal of
  Physics Conference Series} \textbf{118}, 1, 012023 (2008)

\bibitem[{{Reese} et~al.(2017{\natexlab{a}}){Reese}, {Dupret}, \&
  {Rieutord}}]{Reese2017b}
{Reese}, D.~R., {Dupret}, M.-A., {Rieutord}, M., \emph{{Non-adiabatic
  pulsations in ESTER models}}, in Joint TASC2 \& KASC9 Workshop -- SPACEINN \&
  HELAS 8 Conference: Seismology of the Sun and the Distant Stars 2016,
  \emph{European Physical Journal Web of Conferences}, volume 160, 02007
  (2017{\natexlab{a}})

\bibitem[{{Reese} et~al.(2009)}]{Reese2009a}
{Reese}, D.~R., et~al., \emph{{Pulsation modes in rapidly rotating stellar
  models based on the self-consistent field method}}, \emph{A\&A} \textbf{506},
  189 (2009)

\bibitem[{{Reese} et~al.(2013)}]{Reese2013}
{Reese}, D.~R., et~al., \emph{{Mode visibilities in rapidly rotating stars}},
  \emph{A\&A} \textbf{550}, A77 (2013)

\bibitem[{{Reese} et~al.(2017{\natexlab{b}})}]{Reese2017}
{Reese}, D.~R., et~al., \emph{{Frequency regularities of acoustic modes and
  multi-colour mode identification in rapidly rotating stars}}, \emph{A\&A}
  \textbf{601}, A130 (2017{\natexlab{b}})

\bibitem[{{Rieutord} et~al.(2016){Rieutord}, {Espinosa Lara}, \&
  {Putigny}}]{Rieutord2016}
{Rieutord}, M., {Espinosa Lara}, F., {Putigny}, B., \emph{{An algorithm for
  computing the 2D structure of fast rotating stars}}, \emph{Journal of
  Computational Physics} \textbf{318}, 277 (2016)

\bibitem[{{Royer}(2009)}]{Royer2009}
{Royer}, F., \emph{{On the Rotation of A-Type Stars}}, in J.-P. {Rozelot},
  C.~{Neiner} (eds.) The Rotation of Sun and Stars, \emph{Lecture Notes in
  Physics, Berlin Springer Verlag}, volume 765, 207--230 (2009)

\bibitem[{{Townsend}(2003)}]{Townsend2003b}
{Townsend}, R.~H.~D., \emph{{A semi-analytical formula for the light variations
  due to low-frequency g modes in rotating stars}}, \emph{MNRAS} \textbf{343},
  125 (2003)

\bibitem[{{Woosley} \& {Heger}(2006)}]{Woosley2006}
{Woosley}, S.~E., {Heger}, A., \emph{{The Progenitor Stars of Gamma-Ray
  Bursts}}, \emph{ApJ} \textbf{637}, 914 (2006)

\bibitem[{{Zinnecker} \& {Yorke}(2007)}]{Zinnecker2007}
{Zinnecker}, H., {Yorke}, H.~W., \emph{{Toward Understanding Massive Star
  Formation}}, \emph{ARA\&A} \textbf{45}, 481 (2007)

\end{thebibliography}

\end{document}